\documentclass[useAMS,usenatbib]{mn2e}
\usepackage{psfig}
\usepackage{url}

\def\mnras{{\rm MNRAS}}

\def\etal{{\rm et~al.\ }}

\def\kms{{\rm \;km\;s^{-1}}}

\def\simlt{\lower.5ex\hbox{$\; \buildrel < \over \sim \;$}}
\def\simgt{\lower.5ex\hbox{$\; \buildrel > \over \sim \;$}}

\title[Future dark energy constraints]{
Future dark energy constraints from measurements of quasar parallax:
Gaia, SIM and beyond.
}

\author[F. Ding and R.A.C. Croft]{
 Fiona Ding$^{1,2}$ and
 Rupert A.C. Croft$^{1,2}$\\
$^{1}$ Bruce and Astrid McWilliams Center for Cosmology,
Carnegie   Mellon  University,\\
$^{2}$ Dept.   of  Physics,   Carnegie   Mellon  University,
Pittsburgh, PA 15213, USA
}
\begin{document}
\pagerange{\pageref{firstpage}--\pageref{lastpage}} \pubyear{2005}

\maketitle

\label{firstpage}

\begin{abstract}
A consequence of the Earth's well-measured motion with respect to
 the Cosmic Microwave
Background is that over a 10 year period it will travel a distance of 
 $\sim 800$ AU. As first pointed out by Kardashev in 1986,
this distance can be used 
as a baseline to carry out astrometric measurements of quasar parallaxes,
so that only microarcsecond precision is necessary to detect parallax
shifts of objects at gigaparsec distances. Such precision will soon 
be approached with the
launch of the astrometric satellites Gaia and SIM. We use a Fisher 
matrix formalism to investigate
the constraints that these and future, even more ambitious
missions may be able to place on the
cosmological distance scale and the
parameters describing dark energy. { We find that by observing around
a  million quasars as planned, an extended 10 year Gaia mission should have the
capability to detect quasar
parallax shifts at the $2.8 \sigma$ level and so
measure the Hubble constant to within $25 \kms$. For the interferometer SIM,
(in its currently proposed SIMLite configuration) 
a Key Project using $2.4 \%$ of the total mission time to
observe 750 quasars could detect the 
effect at the $ 2 \sigma$ level and dedicated use of the instrument
at the $3.3 \sigma$ level.
In a concordance cosmological model,
Gaia and dedicated SIMLite only weakly
constrain the presence of a cosmological constant
at the $\sim 1 \sigma$ levels.
We also investigate a range of future mission concepts, 
such as an interferometer similar in scope and design to NASA's
Terrestrial Planet Finder. This
could in principle measure the dark energy parameters $w_0$
and $w_a$ with precision $\sigma_{w_{0}}=0.02$ and 
$\sigma_{w_{a}}=0.05$ respectively, yielding a Figure of Merit
larger than the stage IV experiments considered
in the report of the Dark Energy Task Force.}
Unlike perhaps all other probes of dark energy there appear to be
no obvious  astrophysical sources of systematic error on these measurements. 
There is however uncertainty regarding the 
statistical errors. As well as measurement
error, there will be small additional contributions from image
centroiding of variable sources, quasar peculiar 
motions and weak microlensing by stars along the line of sight.

\end{abstract}

\begin{keywords}
Cosmology: observations 
\end{keywords}

\section{Introduction}
The quest to measure the cosmological distance scale has been continued 
over the decades since the discovery of the expansion of the Universe
in many different contexts, from studies of the deceleration parameter, to
more recently dark energy parameters and modified gravity (see e.g., Frieman 
\etal 2008, Jain \& Zhang 2008).
The success of supernova standard candles in revealing the acceleration 
of the Universe (Perlmutter \etal 1999, Riess \etal 1998) 
has shown the power of classical tests,
while at the same time much effort has been and will be spent
in dealing with the many possible systematic errors in the 
measurements. The simplest and
most direct classical test, using pure geometry to measure distances of
objects from their parallax shift over time is arguably the most free 
of systematic uncertainty and
easiest to interpret. More importantly it seems as though carrying out
parallax measurements on cosmological scales should be feasible, through
the combination of astrometric satellites, statistical averaging over many 
objects, and the long baseline afforded by the Earth's motion with respect
to the Cosmic Microwave Background (CMB). 
 In this paper we investigate
how well this combination can be expected to lead to 
cosmic distance scale and dark energy constraints in the future.

The parallax distance to an object in an expanding Universe was first 
calculated theoretically and published by McCrea (1935), although he noted
that it was unlikely to be measurable. We give the
result in the context of dark energy cosmological models in \S2 below.
We note that the calculation was also performed by Kardashev, Pariiskii and
Umarbaeva (1973) who explored the possiblity of measurements using 
radio interferometry. It appears also in the
textbook by Weinberg (1972), and the case of inhomogeneous
universes was treated by Novikov (1977) (and
also Kasai 1988). 

Kardashev (1986) was
the first to propose that the Earth's motion with respect to the CMB
would provide a much longer usable baseline for parallax measurements
and make measurements much less technically challenging than using the
Earth's annual parallax. { This effect is a variant of the ``secular'' 
parallax (see e.g. Binney and Merrifield, 1998, Section 2.2.3), and is in
 principle
easier to measure because the signal increases linearly with time,
while the annual (also known as ``trigonometric'')
parallax repeats at a constant (small) value.}

 Rosquist (1987) pointed out that parallax distances
of distant objects can be used to determine number densities of 
conserved classes of objects such as galaxies even if no dynamical
model is assumed. 
Pierce and Cash (2004) explored the possibility that future X-ray 
interferometers may be able to measure 
the differential parallax between quasar pairs,
and hence characterize dark energy.
Most recently, Quercellini \etal (2008) showed
how alternative anisotropic models such as Lemaitre-Tolman-Bondi cosmologies 
with off-center observers would produce a { secular} 
parallax effect in distant quasars
even for a stationary observer and how upcoming astrometric satellites
may be able to put competitive constraints on those models.

Our plan for this paper is as follows. In \S2 we summarize prior results
for the parallax of extragalactic objects, and generalize them to the case of
time varying dark energy models. In \S3 we give details of planned future
surveys of quasars with astronometric satellites, as well as outlining
some hypothetical, more futuristic surveys. In \S4 we deal with how the 
quasar datasets should be analyzed, and what the systematic and statistical
uncertainties are likely to be. We describe how well these future surveys
can be expected to constrain dark energy parameters in \S5, and in \S6 we
summarise and discuss our results.

\section{Quasar parallax in dark energy models}

The solar system is moving with respect to the CMB frame
at a velocity of $369.5 \pm 3.0$ km/h towards an apex 
with galactic latitude and longitude $l=264.4^{\circ} \pm 0.3^{\circ},
b=48.4^{\circ} \pm 0.5^{\circ}$  (Kogut \etal 1993). As a result,
all
extragalactic objects will experience a parallax shift,
increasing linearly with time, towards the
antapex with amplitude proportional to $\sin\beta$,
where $\beta$ is the angle between the object and the direction of
the apex. Over a 10 
year period a 
baseline of $l=3800\mu \rm{pc}$ is therefore available
for measures of parallax.
We summarize below the expressions for the parallax
shift of a distant extragalactic source (first computed by McCrea, 1935),
using the notation due to Kardashev (1986) and Hogg(1999),
and making them relevant
for a Universe dominated by dark energy.

We write the equation of state for substances in cosmology  as:
\begin{equation}
p=w\rho
\end{equation}
where $p$ is the pressure, $w$ is a dimensionless number, and $\rho$ 
is the energy density. As usual,  dark energy (e.g., Frieman \etal 2008) 
is defined as a substance that has
$w<-\frac{1}{3}$, and thus negative pressure. We follow
the usual parametrization for the equation of state for dark 
energy that varies with time as follows (Chevallier \& Polarski 2001,
Linder 2003):
\begin{equation}
w=w_0+(1-a)w_a,
\end{equation}
where $w_0$ is the value of $w$ at redshift $z=0$, $a$ is
the cosmological scale factor given by $a=1/(1+z)$ and $w_a$ 
governs how $w$ changes with time.

The Hubble parameter in terms of redshift is given by (e.g., Hogg 1999):
\begin{equation}
H(z)=H_{0}\cdot E(z),
\end{equation}
where $H_{0}$ is the Hubble parameter at redshift $z=0$.
 $E(z)$ for dark energy models is given by (Seo and Eisenstein 2003):
\begin{equation} 
E(z)=\sqrt{\Omega_{M}\left(1+z\right)^3+
\Omega_{k}\left(1+z\right)^2+\Omega_{DE}
e^{c(z)}}
\end{equation}
where $w(z)$ is the equation of state for dark energy (Equation 2) and
\begin{equation}
c(z)=3\int^{z}_{0}\frac{1+w(z)}{1+z}dz.
\end{equation}

The equation for the parallax angle is given by:
\begin{equation}
\pi=\frac{l}{r}+\frac{l H_0}{c}
\end{equation}
where $l$ is the baseline, and $r$ is the
parallax distance defined below 
(for a flat Universe, it is equal to the comoving distance).
In the case of a measurement made from a fixed baseline (such as the
usual annual parallax), both terms are relevant. In the case
of a measurement where the baseline is expanding with the Universe (which
is the case for the present paper), only the first term should be used.

The expression for the parallax distance $r$ in Equation 6, depends on 
the curvature of the Universe:
\begin{equation}
r=
\left\{ \begin{array}{ll}
        \frac{c}{H_0} \frac{1}{\sqrt{\Omega_k}} 
\tanh(\frac{\sqrt{\Omega_k}H_0}{c}D) & \Omega_k>0\\
        D & \Omega_k=0\\
        \frac{c}{H_0} \frac{1}{\sqrt{|\Omega_k|}} 
\tan(\frac{\sqrt{|\Omega_k|}H_0}{c}D) & \Omega_k<0
        \end{array},
\right.
\end{equation}
where
\begin{equation}
D=c\int^{z}_{0}\frac{dz'}{H(z)}.
\end{equation}
and $\Omega_k$ is the curvature parameter expressed in terms
of a fraction of the critical energy density.

We note that the expression for the parallax distance is therefore
identical to the angular diameter distance for a flat Universe.
For other cosmologies it differs by the substitution of
tangent for sine and tanh for sinh (see e.g., Hogg 2000 
for the angular diameter distance). For non-flat Universes, therefore,
there is not a direct relationship between the other distance measures
as there is   between  angular diameter  and luminosity distance,
(related by a factor of $(1+z)$). Measurement of parallax distance
therefore promises to reveal new information (see e.g., Rosquist, 1987 for
more discussion) and different parameter degeneracies.

\begin{table}
        \centering
                \begin{tabular} {|c|c|} \hline
                        Redshift & Number of quasars \\ \hline
                        $0.25$ & $49000$ \\
                        $0.75$ & $338240$ \\
                        $1.25$ & $494040$ \\
                        $1.75$ & $407160$ \\
                        $2.25$ & $263650$ \\
                        $2.75$ & $145620$ \\
                        $3.25$ & $57290$ \\
                        $3.75$ & $19020$ \\
                        $4.25$ & $5700$ \\
                        $4.75$ & $1880$ \\
                        $5.25$ & $480$ \\
                        $5.75$ & $90$ \\
                        $6.25$ & $20$ \\ \hline
                \end{tabular}
        \caption{Predicted number of quasars (with SDSS
$i$ band magnitude $\leq 20$) observable over the whole sky, (see \S3)}
        \label{qz}
\end{table}

\begin{figure}
\centerline{
\psfig{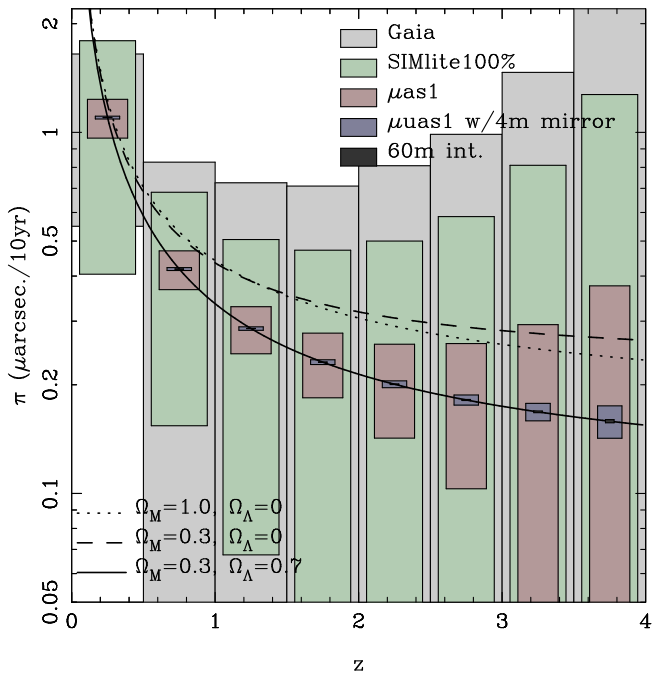}
}
\caption{
The expected { secular} 
parallax shift of quasars over 10 years, as a function of
redshift. We show curves for three different models. Error bars are given
on the $\Omega_{M}=0.3, \Omega_{\Lambda}=0.7$ curve
for 5 of the different experiments from Table \ref{expts}
 { (all except ``SIMLiteKP'').}
The horizontal extent of each error
box is arbitrary.
\label{lines}
}
\end{figure}

\section{Future surveys}

A measurement of quasar { secular}
parallax is not part of the science requirements
for any of the currently planned astrometry missions, 
Gaia (Lindegren \etal 2008),
SIM (Unwin \etal 2008) and Jasmine (Yano \etal 2008),
presumably due to the technical challenges. In the present paper,
we will make predictions for what might be observable with Gaia and SIM, 
in the best possible scenario, but we recognize that for
cosmological constraints to be competitive with other 
methods (at least in terms of statistical errors), significantly more
futuristic instruments will need to be employed.

 We have not made detailed simulations of the performance of any of the
instruments (planned or hypothetical). We have merely used estimates
from the literature of the expected performance of upcoming missions, and 
also crudely extrapolated those to make estimates for 
futuristic { mission concepts.} In any
case, the most important inputs at the level we are doing the
calculations are the number of quasars and the expected astrometric
measurement accuracy per quasar. { Because a secular parallax is being
measured, the relevant error bars are those on the proper motion
of objects, which 
are different to the trigonometric parallax errors. The proper
motion errors  
for  Gaia and SIM
(see below) are quoted in terms of $\mu$arcsec/yr,
for a measurement made over a 5 year period. If the observing period
is increased, the proper motion errors will decrease,
both because the secular parallax increases linearly with time and because
the number of pairs of observations separated by time $\Delta t$ increases.
This yields a proper motion error that scales approximately
as $t^{-3/2}$ (C. Bailer-Jones, private communication),
a factor which we use in our calculations.}
 In this paper, for all the missions, we have assumed a
10 year baseline for observations. In the case of Gaia and SIM, this would
require extensions to the nominal 5 year missions
We note that the current planned extended Gaia mission is 
6 years, (Lindegren \etal. 2008), so that 10 years may be too optimistic.

\subsection{Gaia}

An estimate for the number of quasars that Gaia is expected to observe is
available in the Concept and Technology Study
\footnote{\url{http://www.rssd.esa.int/SA-general/Projects/GAIA_files/LATEX2HTML/report.html}}) (from the year 2000)
carried out by the GAIA Science Advisory Group (D. Schneider,
private communication, data also available online$^{1}$).
We reproduce the expected number of quasars with  Sloan $i$ magnitude
$< 20$ in bins of redshift width $\Delta z=0.5$ in  table \ref{qz}. We
emphasize that although more recent estimates of the luminosity
function have been made since,
at the level of our exploratory analysis the Gaia official predictions are
more than adequate. Finding targets for the astrometry satellites
to observe will rely on prior surveys, such as SDSS (e.g.,
Richards \etal 2009)
For comparison, the number of confirmed quasars with spectroscopically
measured redshifts in the catalog by
Veron-Cetty \& Veron (2006) is  85221.

\begin{table}
{ 
        \centering
                \begin{tabular} {|c|c|c|c|c|c|c|c|c|} \hline
   V mag & 6-13 & 14 & 15 & 16 & 17& 18 & 19& 20 \\ 
   $\sigma_{\rm PM}$ 
($\mu$as yr$^{-1}$) & 5 & 7 &11 &18 &30 &50 &80 &145\\ \hline
                \end{tabular}
        \caption{\label{gaiapm} Sky-averaged rms proper motion error as
a function of visual apparent magnitude for the Gaia satellite (
data from Lindegren \etal 2008, see \S3)}
}
\end{table}

Gaia will observe large numbers of quasars (around a million), 
but the sky survey
will not spend greater periods observing fainter objects, so that the 
positional accuracies will be a function of magnitude. We have used the
magnitudes of the quasars in the  Concept and Technology Study data to compute
for each quasar the $1 \sigma$ expected error on the { secular} parallax,
$\sigma_{\pi}$. { This was done using the proper motion errors from 
Lindegren \etal (2008), which we repeat in table \ref{gaiapm}.
The quasars in the predicted dataset have Sloan $i$ magnitudes between 
12 and 20, so that the faintest have a proper
motion error $\sigma_{\rm PM}=145 \mu$arsec yr$^{-1}$.
To convert this proper motion error into an
esimate of the quasar { secular} parallax measurement uncertainty 
$\sigma_{\pi}$ after 10 years, we use
$\sigma_{\pi}=\sigma_{\rm PM}
\times 10 {\rm yrs} \times (5 {\rm yrs}/10 {\rm yrs})^{1.5}$.
This yields $\sigma_{\pi}=510 \mu$arsec for the faintest quasars, for example
(see Table \ref{expts}).}

We have computed the weighted mean error on the measured parallax for each
redshift bin, assuming Poisson errors, and use these to compute
our Fisher matrix estimates (below). The errors were also used to show
1 $\sigma$ error boxes for a $\Lambda$CDM model
on Figure \ref{lines}. We have multiplied the error bars by
1.26 to account for the fact that the measurements in each redshift bin
will be an average over quasars which each have a parallax
proportional to $\sin \beta$, where $\beta$ is the angle between the
quasar position and the Solar system's direction of motion.

\subsection{SIM}

The Space Interferometry Mission, SIM, also known as SIM PlanetQuest\footnote
{\url{http://planetquest.jpl.nasa.gov/SIM/index.cfm}} is
planned to be a Michelson Interferometer, with 50cm mirrors (there are
4) up to 9m apart. In the current most likely  configuration, known as
SIMlite, the relevant distance will be 6m. It will have
an overall absolute precision of $\sim 3 \mu$arcsec and in narrow
angle mode $\sim 1 \mu$ arcsecs differential accuracy between
objects $<15$ deg. apart.
SIM will not be carrying out a sky survey in the same manner as Gaia,
but will be targeted towards various objectives including extrasolar
planet detection. It would therefore not in principle be a suitable
mission to carry out the current study, given that only of the order of
100 quasars are planned to be observed currently and that measuring
quasar { secular}
 parallax using a larger sample would require dedicating the telescope
to the project at the exclusion of all other science, with (as we
shall show) { limited increases in} possible constraints on cosmology.
A detailed study of the precision
astrometry which will be possible with SIM
 has been published by Unwin \etal (2008). According to this paper,
 approximately $1.5 \%$ of the
mission will be used to measure 50 quasar positions to define the absolute
reference frame. { We do not include those measurements in what 
follows, although we note here that they would definitely be useful
as they will be bright objects, with 
consequently small positional error bars}.
{ Instead, we compute two different scenarios for the use of SIMlite:

The first is a SIMlite Key Project (SIMliteKP in Table \ref{expts}),
with time use guided by the fact
that the 15 currently planned Key Projects$^{2}$ take up $36\%$ of the
observing time. We therefore take a time fraction equal to the average
Key Project fraction, $2.4 \%$ of the mission. For the Key Project,
we also limit ourselves to nearby quasars (redshift $z<0.5$,
this is the only survey for which we do not simply sample
the full $z$ distribution in Table \ref{qz}). Using the SIM time 
estimator\footnote{\url{http://mscws4.ipac.caltech.edu/simtools/portal}}
we find that this would enable 750 16th magnitude quasars to be observed
with proper motion accuracy of 6.0 $\mu$arcsec yr$^{-1}$
(we assume 20 visits per quasar). Although many of
the quasars within $z=0.5$ are brighter than 16th magnitude,
we make the conservative
choice not to use that information. Properly optimizing an
observing strategy is left for future work. 

The second scenario is where $100\%$ of the SIMlite observing time is
used to observe quasars, at the expense of all other science. While this
is not a realistic strategy, it does serve to illustrate what the satellite's
capabilities are. In this case, we find that (with 50 visits per quasar)
9500 quasars can be observed with proper motion 
accuracy of 4.0 $\mu$arcsec yr$^{-1}$ (see SIMLite100\%
in Table \ref{expts}). This scenario, as we sample
the entire quasar $n(z)$ distribution, will not be optimal in terms
of parameter constraints. This should be borne in mind when comparing
constraints from SIMliteKP and SIMlite100\%.

For both of these scenarios, we convert the proper motion error into an
absolute value of $\sigma_{\pi}$ after 10 years in the same fashion as
was done for Gaia (Sec 3.1, above). Because the SIM time estimator
assumes two dimensional results (i.e. errors on $x$ and $y$) 
are required, the proper motion error was reduced by an additional factor 
of $\sqrt{2}$. The resulting $\sigma_{\pi}$ values are given in 
Table \ref{expts}.
}

\subsection{Mission concepts}

We can see from Figure \ref{lines} that the expected error bars on even our 
optimistic constraints from the Gaia and SIMlite satellites are large. A 
detection of quasar { secular} 
parallax from bins with $z<2$ appears possible
 but
constraints on cosmological parameters (we explore these in \S5) will 
be weak. As a result, for precision cosmology it is necessary
to explore constraints from instruments that can provide a 
higher measurement accuracy. We carry out a simple treatment of this
by specifiying the number of quasars to be observed and
the proper motion accuracy per quasar for 3 futuristic satellite
experiments. These are listed as the last 3 entries
in Table \ref{expts}. We emphasize that none of these has potential
missions has been examined in detail and that it is beyond the scope
of this paper to investigate what instruments would be required 
for measurements of quasars
to actually be carried out to the accuracy listed.

 {  The first of these hypothetical experiments (``$\mu$as1'') is meant to
represent a satellite similar to SIM but which can achieve higher accuracy,
on the order of 1 $\mu$ arcsec using a mode like SIM's
``narrow angle mode''. We assume for
$\mu$as1 that 0.8 $\mu$arcsec yr$^{-1}$ proper motion errors are possible
and carry out measurements for 10,000
quasars (this would involve some 17th magnitude quasars). Apart from these,
we assume the same mission parameters as for SIMlite100\% (i.e. a 10 year
mission, using a dedicated satellite). The effective measurement error
on the parallax after 10 years would therefore be 2.0 $\mu$as.}

{ The second,
 (``$\mu$as1 w/4m'') is meant to roughly approximate a version of $\mu$as1 with
4m diameter mirrors (but still with the same baseline (6m)). This would enable
$\mu$as1 w/4m to observe $\sim 100$
times more objects. We note that some of the 
mission time will be allocated to the substantial overhead time associated with
the astrometric data taking. For short integrations ($\sim 30$ secs)
 this can be of the same
 order as the integration time.}
 We also note that the largest mirrors which have
yet been flown in space are those of the Herschel
Observatory (3.5m diameter). 

 The final row is for a mission with similar light 
collecting area, but with an interferometer baseline of 60m, { so that the 
measurement error on the parallax has been reduced to 0.2 $\mu$arcsec over
10 years.} Whether at this level,
the measurement error wil be dominated by other effects is uncertain. We
explore this further in \S4.
{ We note that such a telescope could be similar in scope and design to the
interferometer design for NASA's Terrestrial Planet Finder satellite,
TPF-I\footnote{\url{http://planetquest.jpl.nasa.gov/TPF/}}.
That mission concept has 4 free-flying 4m diameter mirrors in formation
with a maximum 200m baseline. ESA's Darwin 
mission\footnote{\url{http://www.esa.int/science/darwin}}
is another
free-flying optical interferometer of similar size.}

\begin{table}
{
\caption[expts]{\label{expts}
A summary of various surveys discussed in \S3. We give the name
of the experiment, the total number of quasars to be observed, and 
the proper motion measurement error $\sigma_{\rm PM}$
(for the non-extended 5 year 
missions: see e.g., Lindegren \etal 2008 for Gaia). We also list
the resulting effective measurement uncertainty
in the parallax angle $\pi$ per quasar, $\sigma_{\pi}$, after a
10 year observing period (see text). All missions
have all sky coverage, and the quasars are drawn to uniformly
sample the redshift distribution given in table \ref{qz},
except for SIMlite KP for which only quasars with redshift $z<0.5$
are used.
The first three rows have parameters close to currently proposed
experiments, the last three, below the
horizontal line are hypothetical future missions. The last (60m int.)
could be a similar design to NASA's proposed Terrestrial Planet Finder
satellite.}

\begin{tabular}{ccccc}
\hline&\\
Expt.  & $N_{\rm Q}$ &$\sigma_{\rm PM}$ & $\sigma_{\pi}$ & notes \\
  &   &$\mu$as yr$^{-1}$ &$\mu$as &   \\
\hline & \\
Gaia &   1700000 & 5-145 & 18-510  &   \\
SIMlite KP & 750 & 6.0 & 15.0  & 2.4\% time   \\
SIMlite100\% & 9500& 4.0 & 10.0  &   100 \% time  \\
\hline &\\
$\mu$as1  & 10000  & 0.8 & 2.0  &   \\
$\mu$as1 w/4m & 1700000 &0.8 & 2.0  &  4m mirror \\
60m int. & 1700000 & 0.08 & 0.2 &  4m + 60m b.l.\\
\hline&\\
\end{tabular}
}
\end{table}

\section{Analysis and uncertainties}
The analysis of a quasar dataset to extract the parallax distance measurements
promises to be highly involved and be at the limit of the technical
capabilities of the initial experimental setups we have considered. One of the
complications is that quasars themselves are planned to be used by Gaia and
SIM to  define an inertial reference frame. For quasar { secular}
parallax measurements,
quasars obviously cannot be considered to have fixed
angular positions, and so a different approach
will have to be used. Also, in order to constrain dynamical cosmological
models, redshifts will need to be known for all the quasars. As we have only 
made use of information in coarse ($\Delta$z=0.5) bins, these could be
photometric redshifts, although in practice, large surveys such as
BOSS (Schlegel, White and Eisenstein 2009)
will soon generate datasets of hundreds of thousands
of quasars with spectroscopic redshifts. If some sort of redshifts
are known for the
quasars, then the differential { secular} parallax can be measured between
quasars at different redshifts, and if this is done for all pairs of
quasars, the result can be converted into a redshift distance relation.
Exactly how best to do this is beyond the scope of this paper, but we note
that one related, simpler approach which  could be explored 
 would be to consider all quasars 
beyond $z=5$ (for example)
 to define a fixed coordinate system and make measurements
with respect to that. In general, we note that
 although $N_{\rm quasar}^2$ angular
separation measurements will be available, the Poisson
 error on the mean angular
parallax shift  for a given $N_{\rm quasar}$  objects will be $\propto
1/\sqrt{N_{\rm quasar}}$ and not  $1/\sqrt{N_{\rm quasar}^{2}}$.

Although the technical difficulties involved in making a measurement
will be formidable, astrophysical systematic errors
are likely to be subdominant to the effects
of various statistical errors.
The reason for this is that it is difficult to imagine effects
that would cause a systematic shift of quasar angular positions across
the sky in a manner consistent with the { secular} parallax shift.
Because the angular shift we are searching for is in a precisely well
known direction, the various errors are all likely instead to 
add to the random measurement error. One effect
that will however vary across the sky is the shift caused by
aberration due to our motion with respect to the barycenter of the
Milky Way. The motion of the solar system around the center of the Milky
Way causes an aberration of the positions of quasars in the same way that 
the Earth's orbit around the Sun causes stellar aberration. Over time,
with enough precision, this galactocentric acceleration of
the Solar system can be detected from quasar astrometry (see e.g.,
calculations by Kopeikin \& Makarov (2006) who show how SIM
observations of 110 quasars can be  used to do this). This (much larger)
signal would need to be subtracted from the positions of quasars. Its
different directional and redshift dependence to the signal we are looking
for should enable this to be done cleanly.

One source of random error will be quasar peculiar motion. This can be divided
into two components, motion of the host galaxy, and the potentially large
orbital motion of black hole binaries as they merge. The effect of
the first of these can be considered by assuming that the magnitude of
the rms motion of
galaxies wrt to the CMB will be similar to our own motion. By elementary
trigonometry, such velocities
will cause galaxies to have angular proper motions from 
our point of view that are similar in amplitude to the { secular} 
parallax shift
due to our motion. The crucial difference is that these proper motions
will  be in random directions, and therefore their effect will be negligible
once we average over several thousand (or tens of thousands) of quasars. We
note that the overall
cosmic variance limit for quasar parallax meaasurements
of the distance scale will be due to either
the effects of these peculiar  velocities (which will also affect the redshifts
we use) or else due to 
a bound on the number of faint magnitude quasars. 
{ We note that quasar peculiar velocities will not be
totally random, due to velocity correlations that arise 
during the process of large-scale
structure formation. Because quasars form an extremely sparse sampling
of the cosmic density field, this effect will not be as important
as it would be for galaxies, for example. We leave
determination of the effect of velocity correlations to future work.}

Consideration of the
 second peculiar motion effect, of binary quasars is motivated by
merger models of quasar formation (e.g., Di Matteo \etal 2005), and observations
which show an excess in the number of quasar pairs at small separations
(e.g., Hennawi \etal 2006). For these velocities to disrupt
the parallax distance measurements, however, they would have to be extremely
large, for example in a bin where we average over 10000 quasars,
the rms motion of all quasars
would have to be greater than $\sqrt{10000}$ times our
motion wrt the CMB, or $\simgt 30000 \kms$ to dominate the 
overall error budget. Relativistic jets could also contribute
a large proper motion, although they will only be relevant for 
a small fraction of quasars.
Plots of the quasar autocorrelation function (e.g., Hoyle \etal 2002) show no
sign of extreme elongation in the line-of-sight direction, and so such large
peculiar motions cannot be common for quasars in any case.

Other sources of error include microlensing by stars in the Milky Way and
in the quasar host galaxies. Belokurov and Evans (2002) have estimated 
for Gaia that the all-sky
astrometric microlensing optical depth is  $2.5 \times 10^{-5}$, i.e.
less than one in 10000 sources will have a significant centroid shift.
 Smaller  centroid shifts will however be caused by so-called
weak microlensing, which can also cause negative parallaxes (Sazhin et al. 
2001). Sazhin \etal show that due to weak microlensing, all quasars
are expected to show negative parallaxes on the order of a few
nanoarcseconds, with a small fraction, of the order of 1 $\%$ 
reaching 1 microarcsecond. Apart from lensing errors, there
will be image centroiding
problems due to the underlying quasar host galaxies (see Bastian and Hefele
2004 for a summary). In order to make a convincing detection of
quasar { secular} 
parallax, all of these will need to be dealt with. As for the 
measurement averages will be made over a large number of quasars, the hope
is that these statistical errors will all average out, with no residual
bias. The technical measurement challenges will of course be considerable.
We note that the Gaia Concept and Technology Study has drawn
the pessimistic conclusion that these effects will average
between $10\mu$ arcsec and $100 \mu$ arcsec for a typical quasar, much
larger than we assume here (where for example our 60 m interferometer
predictions assume a total error per quasar of $0.2 \mu$ arcsec.) This
is at variance with other predictions (e.g., Bastian and Hefele
2004),  and the hope is that further research (and Gaia observations
themselves of nearby quasars) will show such large effects
to be nonexistent.

\section{Predicted constraints from future observations}

We use the survey parameters described in \S3 and summarized in 
Table \ref{constraints} to compute predicted constraints on cosmological
parameters from future observations of
quasar { secular} parallax.

In each case, we use as the underlying model the $\Lambda$CDM model
plotted as a solid line in Figure \ref{lines} (i.e. $\Omega_{\Lambda}=0.7$,
$\Omega_{M}=0.3$, h=0.72, w$_{0}$=-1, w$_{a}$=0. We compute
estimated $1\sigma$ uncertainties for the bins of redshift width
$\Delta$ z=0.5 using Poisson statistics.

For the cosmological 
parameter constraints themselves, we use the
standard Fisher matrix approach (see e.g., Tegmark \etal 1997).

The Fisher Matrix is given by:
\begin{equation}
F_{jk}=\sum_{b}\frac{1}{\sigma^{2}_{b}}\frac{\partial f_b}{\partial p_j}
\frac{\partial f_b}{\partial p_k}
\label{eqn:Fisher}
\end{equation}
where $f_b$ is the function in terms of the bin $b$, $p_j$ and $p_k$ 
are parameters of $f$, and 
$\sigma^2_b$ is the variance in $b$.

A Gaussian prior on any of the parameters can be added by adding
$(\sigma_{\rm prior}^{2})$ to the appropriate diagonal element of the
$F$.  
Inversion of $F$ yields 
the covariance matrix, the diagonal elements of which are 1 $\sigma$
errors on the parameters. To compute the error ellipses, we use the fact 
that the square roots of
the eigen values of the covariance matrix correspond to the error ellipse
axis lengths, and their associated eigenvectors 
define the error ellipse axis directions. The $95\%$ probability ellipse
has axis lengths $\sqrt{6.17}$ times longer than the $68 \%$ ellipse.

\begin{table}
\caption[constraints]{\label{constraints}
Constraints on cosmological parameters from the different experiments
considered in this paper (see Table \ref{expts}, the horizontal
line below divides planned missions from hypothetical future missions).
 For the first 
column (constraints on $H_{0}$), priors on $\Omega_{\Lambda}$
 and $\Omega_{M}$ of 10\% were
assumed. For the second column, a 10\% prior on $\Omega_{M}$ 
was assumed. For the last 3 columns, a 10\% prior on $\Omega_{M}$
and a flat 
Universe were assumed. The column labelled FOM corresponds to 
the Dark Energy Task Force Figure of Merit (see \S5.3).
}
{
\begin{tabular}{cccccc}
\hline&\\
Expt.  & $\sigma_{H_{0}}$ & $\sigma_{\Omega_{\Lambda}}$ & $\sigma_{w_{0}}$
  & $\sigma_{w_{a}}$ & FOM \\
  & (kms$^{-1}$  &  &   & &\\
  &  &  \\
  & Mpc$^{-1}$  &  &   & &\\
\hline & \\
Gaia & 26.4 & 1.06  & 6.7  & 44.5  & 0.0004  \\
SIMliteKP & 35.9 & 3.8  & -  & - & -   \\
SIMlite100\% & 22.2 & 0.58  & 6.8  & 37.3 & 0.0008   \\
\hline & \\
$\mu$as1  & 8.2  & 0.114  & 1.36 & 7.28 & 0.015 \\
$\mu$as1 w/4m & 0.72 & 0.0090  &  0.19   & 0.54 & 0.55\\
60m int. & 0.08 & 0.0009  &  0.019   & 0.054  & 55\\
\hline&\\
\end{tabular}
}
\end{table}

\subsection{Constraints on the Hubble constant}

After a detection of { secular} parallax, 
the first constraints that could be usefully
placed are those on the Hubble constant. This measure of $H_{0}$ would be
based on large scale expansion of the Universe, on gigaparsec scales,
and so would not suffer from peculiar velocity scatter which forms 
part of the error budget in the HST cepheid-based measurements of the distance
scale (Freedman \etal 2001).
Because the Universe has evolved over the light travel
time to these high redshift quasars, it is necessary to assume a prior
on the cosmological model to constrain $H_{0}$. We assume that the
underlying model is a $\Lambda$ CDM model and that
$\Omega_\Lambda$ and $\Omega_M$ are each known to $10 \%$.

The corresponding 1 $\sigma$ error bars on $H_{0}$ for the
different experiments from Table \ref{expts} are shown in Table 
\ref{constraints}.
{ We can see that Gaia measurements, for example could yield a $\sim 35\%$
error bar on  $H_{0}$. While this is not competitive with for example
the constraints from HST key project (Freedman \etal 2001)
it is within a factor of 3. We note also that the Gaia result amounts to
a $2.8 \sigma$ detection of quasar parallax.
The SIMLiteKP
constraint is somewhat worse, but still amounts to 
a 2 $\sigma$ detection. We note that the SIMLiteKP constraint is only a factor 
of 1.6 weaker than the SIMLite100\% constraint, despite using only 2.4\%
of the observing time. This is because we restricted the redshift range
of quasar targets to $z<0.5$, where the parallax shift is largest.
This is partly because there are sufficient $m_{V}\leq 16$ quasars in that
redshift range to do this for SIMLiteKP. The other reason
is that we decided in the SIMLiteKP
to trade detection significance with constraints on dark energy
parameters, which are sensitive to quasars at higher $z$.

 In these calculations,
a 10 year duration has been assumed (\S3).
 If this is instead reduced to 5 years,
the significance of detections for both Gaia and
SIMliteKP shrinks to below $1 \sigma$. 
  Moving through the other, hypothetical
experiments, we can see the error bar on $H_{0}$ rapidly
becomes smaller, as would be expected.

 Other versions of the parallax can extend geometric distance measurement
out as far as local group galaxies. For example, the ``Rotational Parallax''
technique which compares measurements of proper motions
and galaxy rotation curves could be used to estimate bias free distances to
M33, M31 and the LMC to the percent level (Olling 2007). Olling (2007)
has also shown how important such local anchoring of the $H_{0}$
distance ladder is to measurements of other cosmological parameters.}

We note that detection of quasar parallax at the level
predicted would be incontrovertible 
evidence that quasar redshifts are cosmological (c.f.
Burbidge \etal 2003).

\subsection{Cosmological constant dominated models}
The Type IA supernova measurements of the accelerating Universe were initally
used to constrain the parameters  $\Omega_\Lambda$ and $\Omega_M$ (e.g.
Permutter \etal 1999). 
Measurements of quasar parallax could also be used to do this, it being 
interesting to compare the kind of constraints that the first generation
of quasar parallax measurements could make with the first measurements
from supernovae. We have first computed constraints on  $\Omega_\Lambda$ 
for the different experiments, assuming a prior 
on  $\Omega_M$ of $10\%$ (i.e. $\sigma_{\Omega_M}=0.03$).
The results are extremely insensitive to the size of the prior.
The $1\sigma$ errors on  $\Omega_\Lambda$ are given in Table
\ref{constraints}, { where we can see that Gaia and SIMLite100\%
can only in principle achieve
very weak  $0.7 \sigma$ and $1.3 \sigma$ constraints on $\Lambda$
respectively. To constrain dark energy parameters therefore we will need
to wait for future mission concepts to become reality.
These other survey concepts could do significantly better,} 
with extremely tight constraints possible from the more futuristic ones
(at the level of $0.07 \%$ for 60m int. experiment). 

\begin{figure}
\centerline{
\psfig{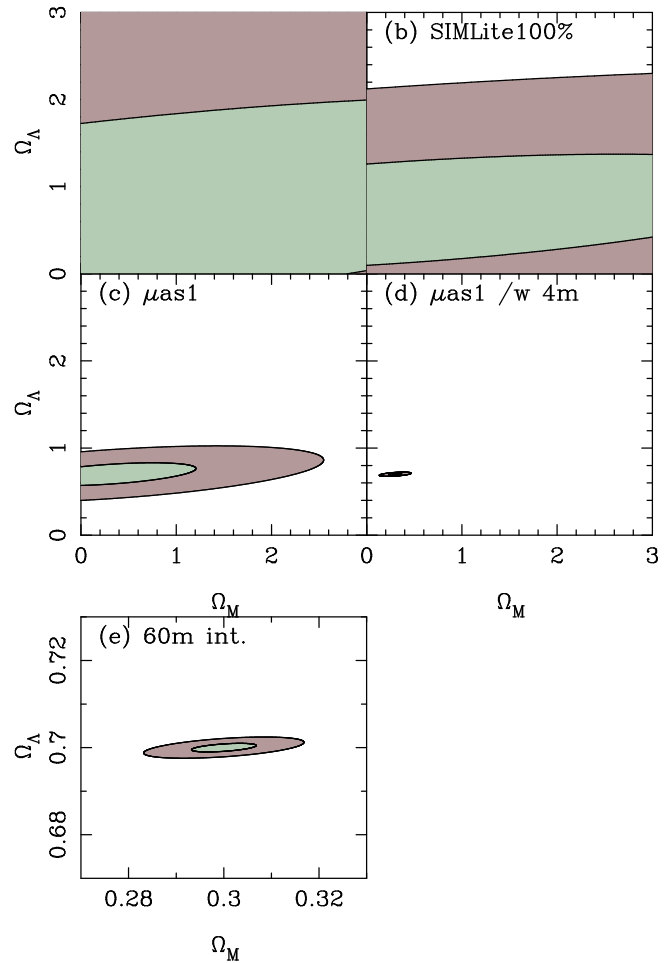}
}
\caption{
Future possible constraints on the parameters
$\Omega_{M}, \Omega_{\Lambda}$ (with no prior on $H_{0}$) calculated
using the Fisher matrix formalism. 
We show $68\%$ and $95\%$ confidence ellipses
for 5 different experiments (the same as in Figure \ref{lines}).
Note the change in axis scale for panel (e).
\label{omlam}
}
\end{figure}

The reason for the relative tightness of the constraints becomes apparent
once we plot the Fisher matrix contours for  $\Omega_\Lambda$ and $\Omega_M$
(Figure \ref{omlam}), where we can see that the contours are quite close to 
horizontal, i.e. quasar parallax measurements constrain $\Omega_\Lambda$
well but are unable to say much about  $\Omega_M$. The difference between
the expression for the  parallax distance 
(Equation 7)
and the luminosity distance (or angular 
diameter distance which is just related by (1+z) factors)) causes
a difference in the direction of least degeneracy.  SN measurements
(e.g., Riess \etal 1998)
constrain $\Omega_\Lambda$ and $\Omega_M$ approximately equally.
Measurements of galaxy cluster abundances (e.g.,  Allen \etal 2002) on the 
other hand constrain $\Omega_M$ much better. 

\subsection{Redshift-varying dark energy models} 

\begin{figure}
\centerline{
\psfig{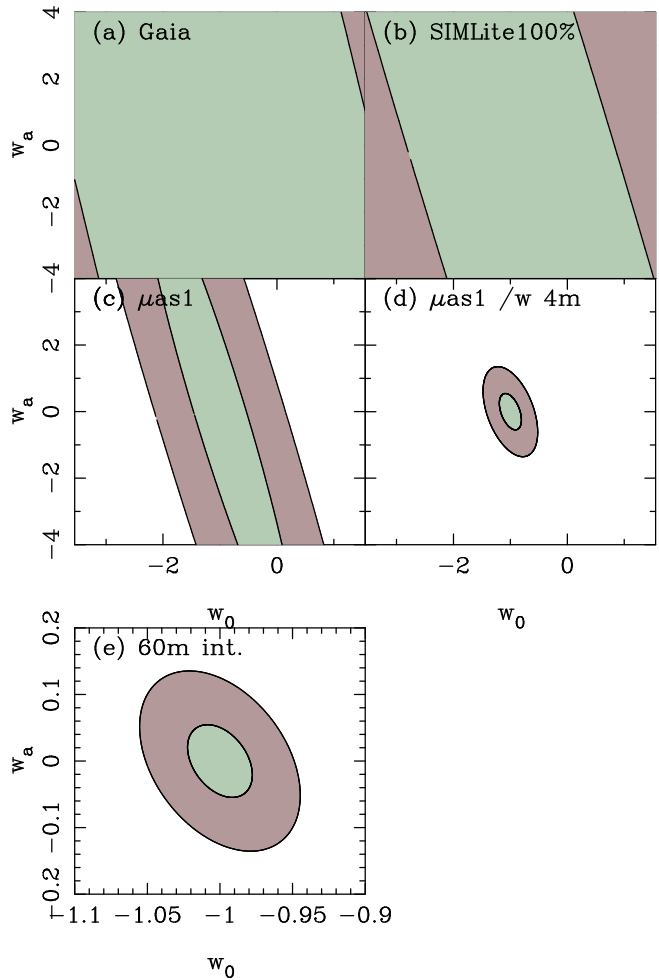}
}
\caption{
Future possible constraints on the parameters
$w_{0}, w_{a}$  calculated
using the Fisher matrix formalism. 
We show $68\%$ and $95\%$ confidence ellipses
for 5 different experiments (the same as in Figure \ref{lines}).
Note the change in axis scale for panel (e).
\label{wa}
}
\end{figure}

The nature of dark energy itself is not probed by a precise determination
of $\Omega_\Lambda$. To do that, the Dark Energy Task Force (
DETF, Albrecht
\etal 2006) has 
recommended that the variation of dark energy with redshift be determined,
using the parametrization given in Equation 2. We have computed the constraints
on $w_0$ and $w_a$, with our standard prior on  $\Omega_M$ of $10 \%$. We
have also assumed a flat model in this calculation, although we describe the
effect of relaxing this assumption below. The $1 \sigma$ constraints
on the parameters are listed in Table \ref{constraints}, and the 
confidence contours are plotted in Figure \ref{wa} for the various
experiments. We can see that as is usual with other probes of dark energy,
$w_{a}$ is significantly more loosely bounded than $w_{0}$. For experiments
smaller than { $\mu$as1 w/4m} there is little useful constraint on $w_{a}$.
For the extremely ambitious experiment 60m int., measurement
of $w_0$ at the $1\%$ level would be possible and a 0.03  value of $w_{a}$
would be detectable.  Without the flat space prior, we have calculated
that these would increase to $\sigma_{w_{0}}=0.032$ and $\sigma_{w_{a}}=0.12$.

We have computed the inverse of the area enclosed by the
 $95\%$ confidence contours, to be compared to the DETF Figure of Merit
(FOM). These values are also shown in Table \ref{constraints}.
In the Table of results (page 77) in the DETF report, values are
tabulated not for the FOM, but for a quantity proportional to it,
$1/(\sigma_{w_a}\sigma_{w_p})$, where $w_p$ is the value of
$w$ at a pivot redshift chosen
so that the errors in $w_{a}$ and $w_{p}$ are uncorrelated.
They are related by $FOM/6.17\pi=1/(\sigma_{w_a}\sigma_{w_p})$,
so that{ $1/(\sigma_{w_a}\sigma_{w_p})=0.008, 0.016, 0.29, 10.7, 1070$}
for Gaia through 60m int.
Comparing these values to those in the DETF report, we can see that
only the extreme experiments, { $\mu$as1 w/4m} and 60m. int are competitive
with so called stage III and stage IV surveys. { The 60m. int is
more constraining than all the DETF survey models (the
highest FOM considered by the DETF is weak lensing
with the Square Kilometer Array [``optimistic case''],
$1/(\sigma_{w_a}\sigma_{w_p})=645.76$,
a factor of 1.7 times smaller than for 60m. int).
The { $\mu$as1 w/4m} case is as good as several
of the listed stage IV surveys.
}

In carrying out these comparisons, we note that the DETF report does not
assume a flat space prior, and also that the important results for the
DETF are how the constraints can be combined with other measurements
(for example from stage II surveys) to increase the FOM. Without a flat
space prior, the constraints from the parallax experiments become worse,
{ but they are still competitive: $1/(\sigma_{w_a}\sigma_{w_p})=9.0$
and $605$, respectively
for $mu$as1 w/4m and 60m int., the latter still comparable
to the best case studied by the DETF.}
As for the combination
of constraints with those from other methods, although it is beyond the
scope of the present paper to investigate this, it is clear from
the equation for the parallax distance (Equation 7) that since they are not
related to the angular diameter and luminosity distance by simple factors
of (1+z) there should be scope for using their different 
degeneracy directions to get good combined constraints. In general though,
because in principle the parallax constraints have the potential 
for less systematic errors than other methods it would be beneficial to 
have small errors on parameters for parallax alone.

We note that as pointed out by the DETF the FOM for the different experiments 
should scale as the number of
quasars observed for equal precision in each
measurement.

\section{Summary and Discusssion}

\subsection{Summary}
In this paper we have briefly explored the potential of astrometric
measurements of quasar { secular} 
parallax as a probe of dark energy in the Universe.
Our conclusions can be summarized as follows:

(1) { Both the currently planned astrometric 
satellite Gaia (due for launch 
by ESA in early 2012) 
and NASA's SIMlite mission (no launch date yet) could in principle detect 
(at the 2-3 $\sigma$ level) the 
quasar secular parallax shift due to the solar system motion with
 respect to the CMB
over extended mission lifetimes of 10 years.}

(2) 
{ Gaia may be able to constrain the Hubble constant at the
 $\sim 25 \kms$ level
but only very weakly 
constrain the acceleration of the Universe (at the $0.7 \sigma$ level).
SIMlite would also be able to offer weak constraints at the $1.3 \sigma$
level on the acceleration but only in the unrealistic case of using 100\%
of the satellite's observing time.
}

(3) { Futuristic astrometric survey instruments, such as 
a large scale free flying interferometer similar in 
size to NASA's Terrestrial Planet Finder could in principle carry out 
precision cosmology. Such a mission could achieve
constraints on  the dark energy
parameters $w_{0}$ and $w_{a}$ significantly  better than  the best of the
``Stage IV'' surveys considered by the Dark Energy Task Force. }

(4) { Less ambitious mission concepts, such as an interferometer
with the same baseline as SIM but with 4m diameter mirrors 
are approximately in the same class in terms of both mirror size and 
dark energy constraining power as some candidates proposed
for the NASA/DOE Joint Dark Energy Mission}

(5) Although the astrophysical systematic errors in the method
seem likely be non-existent, the measurement errors and other statistical
errors are significantly uncertain and their investigation
requires much future work.

\subsection{Discussion}

Although the ultimate aim of the kind of work set out in this paper is
ambitious: precision measurement of dark energy parameters, the 
immediate aim of our work is merely to draw attention to the fact that
upcoming interferometers may actually make direct geometric measurement of
gigaparsec distances possible. Making use of
this exciting  possiblity to  do cosmology
is probably as feasible as other suggested
measurements in ``real time'' cosmology, such as the Sandage-Loeb
test (Sandage 1962, Loeb 1998).
In that probe,  the Universe's expansion rate is
probed directly by looking at the variation of the redshifts of extragalactic
objects with time (say a 10 year timeline, similar to the one we
assume in this paper for our real time parallax shift measurements).
Common to both approaches is the need for immense control of measurement
error and the design of extremely futuristic instruments 
(e.g., Liske \etal 2008) to achieve the goals.
Whether such a vast undertaking should be motivated only by  the
task of measuring dark energy parameters is debatable (e.g., White 2007),
but of course other equally exciting problems require precision
astrometry (e.g., Schneider \etal 2008).

The main advantage of a pure geometric method for constraining dark
energy is the apparent lack of astrophysical systematic effects. This
said, however, it is important to devise tests for unknown and unthought
of errors. As the parallax shift is with respect to a well defined
direction on the sky, this should be straightforward in principle.
For example, perhaps the simplest null test one could imagine would involve
recomputing the parallax shifts assuming that the solar system direction 
of motion is perpendicular to its actual motion, expecting to find
a result of zero within the statistical errors. The statistical 
errors are perhaps the largest source of uncertainty in whether the
cosmological
measurements will be feasible. Precise modelling of image centroiding of
variable sources, and actual Gaia observations will be needed to 
determine how much these will affect measurements. Quasar peculiar
motions will need to be investigated, for example for nearby objects
by Gaia and SIM, for which the peculiar motions should be readily
detectable if they are large enough to be a problem for
the cosmological constraints in this paper. As we have mentioned,
there are some reasons to believe that statistical errors
will be subdominant to the measurement errors, but without significant
future work we cannot be sure.

One of the topics we have not explored here is the optimal 
redshift distribution of quasars needed for the measurement. Of
course this depends on the type of dark energy being tested, but
for the standard parametrization, it is possible that observing time
would be wasted looking at very high redshift quasars, with little
gain in parameter constraints. On the other hand, due to the high
redshifts themselves, quasar distance measurements could perhaps be used
to constrain exotic models of early dark energy better than lower $z$
supernova measurements.

{ We have also not explored how Gaia and SIM results could be
combined to yield stronger constraints on cosmology. The general 
question of how to optimize use available resources to give the
best possible constraints is one which should be considered.
A related interesting question is whether longer timelines
could be used to good effect, particularly in combining Gaia and SIM
with missions which might come later. Also, if Gaia is launched first,
SIM can be used to observe the brighter quasars later and so make use
of a longer timeline.}

What some have called the upcoming ``age of astrometry'' has
the potential to yield the most direct constraints on cosmological
models possible. Achieving precision measures of dark energy
from quasar parallax will require overcoming enormous technical
challenges, but given the attention focussed on many other probes of
dark energy, many of which have astrophysical systematics which
will be very difficult to control, it is worth exploring the matter further.

\section*{Acknowledgments}
We thank the anonymous referee for suggesting many improvements
and corrections to
the paper.
RACC thanks Rachel Bean, Douglas Clowe,
Bhuvnesh Jain, Nikhil Padmanabhan and Jeff Peterson for useful discussions
and Coryn Bailer-Jones for comments which improved the manuscript.

{}

\end{document}